\documentclass[twocolumn, superscriptaddress,nofootinbib, floatfix, amsfonts, aps]{revtex4}
\usepackage{amsmath}
\usepackage{bm}
\usepackage{graphicx}
\usepackage{mathrsfs}
\usepackage{footmisc}
\usepackage{multirow}
\usepackage{graphicx}
\usepackage{booktabs, ctable}
\usepackage{soul,xcolor}
\usepackage{xcolor, ulem, color, framed}
\usepackage{amssymb}

\begin{document}

\title{Charmonium dissociation at high baryon chemical potential}

\author{Bo Tong}
\affiliation{Department of Physics, Tianjin University, Tianjin 300354, China}
\author{Baoyi Chen}\email{baoyi.chen@tju.edu.cn}
\affiliation{Department of Physics, Tianjin University, Tianjin 300354, China}

\date{\today}

\begin{abstract}
We study the charmonium dissociation in the hot medium with 
finite baryon chemical potential $\mu_B$. 
Charmonium bound states 
are dissociated in the medium by the color screening effect and 
the random scatterings with thermal partons, {which are included in the real and imaginary parts of the potential respectively}.  
$J/\psi$ fraction in the $c\bar c$ pair 
defined to be the quantum overlap between the wave package and the wave function of 
$J/\psi$ eigenstate decreases with time due to the complex 
potentials. 
When $\mu_B$ is large compared with the medium temperature, {the Deybe mass is increased evidently}.
{We consider $\mu_B$-dependent Deybe mass in  both real and imaginary parts of the potential} to 
calculate the $J/\psi$ survival probability in the static 
medium 
and the Bjorken medium. 
$J/\psi$ survival probability is reduced evidently 
by the $\mu_B$ effect at low 
temperatures available in the medium produced in Beam Energy Scan experiments, while this effect becomes not apparent at high temperatures.

\end{abstract}
\pacs{ }
\maketitle

\section{Introduction}

Hot deconfined medium is believed to be produced in the relativistic heavy-ion collisions~\cite{Aoki:2006we,Bazavov:2011nk}. 
Heavy quarkonium has been extensively studied to extract the properties of the hot QCD matter in nuclear collisions~\cite{Matsui:1986dk,Grandchamp:2001pf,Andronic:2003zv,Yan:2006ve,Liu:2010ej,Chen:2013wmr,Chen:2019qzx,Zhao:2020jqu,Rothkopf:2019ipj}. 
In the hot medium, heavy quark potential is color screened by the thermal partons~\cite{Karsch:2005nk,Satz:2005hx,Burnier:2015tda}, 
which can dissociate the bound states of quarkonium. The degree of color screening depends on the densities of thermal partons represented by the medium temperature.  
With the increase of the temperature, different quarkonium bound states are sequentially melted  
due to their different binding energies. Besides, inelastic random scatterings from the thermal partons can also dissociate quarkonium bound states~\cite{Peskin:1979va,Burnier:2016mxc,Lafferty:2019jpr,Chen:2018kfo,Zhao:2021voa,Blaizot:2015hya}, 
where heavy quark pair is transformed from singlet to octet states. The singlet-octet transition process 
can be treated as an imaginary potential
which reduces 
the normalization of the singlet states~\cite{Burnier:2016mxc,Krouppa:2015yoa,Boyd:2019arx}. 
One can determine the medium temperature 
with charmonium survival probability defined as the ratio of 
final and initial production of $J/\psi$ during their evolution in the hot medium. 
Explicit quantum treatments have been developed to study the quarkonium inner evolution 
in the medium, such as the Schr\"odinger-Langevin equation~\cite{Katz:2015qja},  which evolves the wave function of the quarkonium directly. The medium interaction is included via the screened potential and the noise term in the Hamiltonian. 
The Schr\"odinger equation model 
with complex potentials are also developed~\cite{Kajimoto:2017rel,Islam:2020bnp,Wen:2022utn}. {The inner 
evolutions of the quarkonium are described with the Schr\"odinger equation when they move along 
different trajectories in the medium.}  
Open quantum system models such as 
the Lindblad equation~\cite{Brambilla:2020qwo} and the Stochastic Schr\"odinger equation~\cite{Akamatsu:2011se,Xie:2022tzs} are also developed recently which treats quarkonium as an open quantum system with the momentum-energy exchange with the thermal medium. Other semi-classical transport models are also developed to study the dissociation and recombination of quarkonium in the hot 
medium~\cite{Yao:2018sgn,Yao:2018nmy,Yao:2020eqy}.

At the experiments of the Beam Energy Scan (BES), 
the initial energy density of the medium is much lower than the situation in AA collisions at 
the Relativistic Heavy-Ion Collider (RHIC) 
and the Large Hadron Collider (LHC). The effects of color screening and the parton random collisions become weaker in the heavy quark potential. 
However, the baryon chemical potential $\mu_B$ in the 
medium produced in BES can be considerable. It changes 
the Deybe mass and the heavy quark potential~\cite{Doring:2005ih,Kakade:2015laa}.  
{It is necessary to study the $\mu_B$-effect on charmonium evolution in the baryon-rich medium with a low temperature and a large $\mu_B$.} In this work, we employ the time-dependent Schr\"odinger 
equation with the complex potential to study the evolution of charmonium wave function in the medium with 
high baryon chemical potential~\cite{Liu:2020cqa,Wen:2022utn}. 
The Deybe mass becomes larger due to the correction from $\mu_B$ term. This results in a weaker real potential and a larger imaginary potential of the quarkonium. 
{$J/\psi$ fraction in the charm pair is more reduced after considering the $\mu_B$ effect in the static and the Bjorken medium. 
Studying charmonium dissociation in high $\mu_B$ medium helps to understand the charmonium evolutions at the experiments of the BES. }

This work is organized as follows. In Section II, we introduce the framework of the Schr\"odinger equation and the parametrized in-medium heavy quark potential. In Section III, the 
evolutions of charmonium wave package in the static medium and the bjorken medium are 
studied respectively. Effects of the baryon 
chemical potential and the color screening are compared in the charmonium dissociations. 
In Section IV, a conclusion is given.

\section{Theoretical model}
To describe the quantum evolutions of heavy quarkonium wave packages at finite $\mu_B$ and $T$, we employ the 
time-dependent Schr\"odinger equation. Neglect the 
relativistic effect in the inner motion of 
charmonium, we take the classical form of the Hamiltonian of charmonium.  
Hot medium effects are included via the 
in-medium heavy quark potential. 
As the QCD matter produced in heavy-ion collisions 
is close to a perfect liquid with very small viscosity, 
one can approximate the heavy quark potential to be a 
spherically symmetric potential. There is no 
transitions between the states with different 
angular momentum. We separate the radial part 
of the Schr\"odinger equation in the center of 
mass frame~\cite{Wen:2022utn}, 

\begin{align}
\label{fun-rad-sch}
i\hbar {\partial \over \partial t}\psi( r, t) = [-{\hbar^2\over 2m_\mu}{\partial ^2\over \partial r^2} +V( r,T)+{l(l+1)\hbar^2\over 2m_\mu r^2}]\psi(r,t)
\end{align}
where $r$ and $t$ are the radius and the time 
respectively. $m_\mu=m_1m_2/(m_1+m_2)=m_c/2$ is the reduced mass in the center of mass frame. $m_c$ is the charm quark mass. 
Heavy quark potential $V(r,T)$ 
depends on the temperature and the radius, which indicates that different eigenstates in the wave 
package experience different hot medium effects due 
to their geometry sizes. 
$\psi(r,t)\equiv r R(r,t)$ is defined as 
the product of the radius and the radial part 
of the wave package $R(r,t)$. The total 
wave package of heavy quarkonium 
is expanded as ${\Psi(r,\theta, \phi, t)}=\sum_{nlm}c_{nl}(t) R_{nl}(r,t)Y_{lm}(\theta,\phi)$. $Y_{lm}$ is the 
spherical function. $(n,l,m)$ are the quantum numbers of 
charmonium states. 
The coefficient $c_{nl}(t)$ is defined to be, 
\begin{align}
\label{eq-project}
c_{nl}(t) &=  \int R_{nl}(r)e^{-iE_{nl} t} \psi(r,t) rdr 
\end{align}
where 
$|c_{nl}|^2$ is interpreted as 
the fraction of the charmonium eigenstate specified with the quantum number $(n,l)$ in the total wave package. The charmonium eigenstates 
mentioned in this work is defined as the eigenstates 
of the vacuum Cornell potential with a string breaking at $r=r_{D\bar D}$,
\begin{align}
 V_{\mathrm{c}}(r)=\left\{
\begin{array}{rcl}
\label{fun-cornell}
-{\alpha\over r}+\sigma r       &      & {r     <      r_{D\bar D}}\\
2m_{D}-2m_c     &      & {r \ge r_{D\bar D}}\\
\end{array} \right. 
\end{align}
where the distance of string breaking $r_{D\bar D}$ 
is determined via 
$-{\alpha\over r_{D\bar D}}+\sigma r_{D\bar D} = 2m_D -2m_c$. Masses of D meson and charm quark is taken 
as $m_D=1.87$ GeV and $m_c$=1.27 GeV~\cite{ParticleDataGroup:2018ovx} respectively. 
Fitting the 
masses of $J/\psi$ and $\psi(2s)$ given by 
particle data group, one can determine the 
values of the parameters $\alpha=\pi/12$ and 
$\sigma=0.2\ \rm{GeV^2}$~\cite{Satz:2005hx}.
With the in-medium heavy quark potential $V(r,T)$, fractions of charmonium eigenstates in the wave 
package change with time. The survival probability 
of charmonium eigenstates is connected with 
the evolutions of charmonium wave package. 
The quantum transition between 
different states have been included in the 
wave function evolutions.

To solve the Schr\"odinger equation numerically, 
we employ the Crank-Nicolson method. It can evolve the 
wave package straight-forward in the spatial coordiante instead of projecting the wave package to a series of basis. The numerical errors of the wave function at different time steps is small enough and convergent when we take a small step of time and the 
radius in the discrete formula (in natural units $\hbar=c=1$),
\begin{align}
  \label{eq-sch-num}
{\bf T}_{j,k}^{n+1}\psi_{k}^{n+1} = \mathcal{V}_{j}^{n}.
\end{align}
where $j$ and $k$ are the indexes of rows and columns 
in the triangular matrix $\bf T$. 
The non-zero elements in the matrix are, 
\begin{align}
\label{eq-sim-cn}
&{\bf T}^{n+1}_{j,j}= 2+2a+b\mathcal{V}_j^{n+1}, \nonumber \\
&{\bf T}^{n+1}_{j,j+1}={\bf T}^{n+1}_{j+1,j}= -a, \nonumber \\
&\mathcal{V}_j^n= a\psi_{j-1}^n +(2-2a-bV_j^n)\psi_j^n +a\psi_{j+1}^n ,
\end{align}
where $a= i\Delta t/(2m_\mu (\Delta r)^2)$ and $b=i\Delta t$. Here $i$ is an imaginary unit. 
The subscript $j$ and superscript $n$ represent 
the coordinate $r_j=j\cdot \Delta r$
and $t_n=n\cdot \Delta t$ respectively. 
The steps of the radius and the time are taken 
to be $\Delta r=0.03$ fm and $\Delta t=0.001$ fm/c. The 
numerical accuracy in the evolution of the wave 
package is high enough when taking these parameters.  
The time dependence in the potential comes from 
the time evolution of the temperature. In the 
static medium with a constant temperature, the potential does not depend on time anymore. At 
each time step, we calculate the inverse of the 
matrix $\bf T$ with the “Gauss-Jordan element elimination” method in Eq.(\ref{eq-sch-num}) to obtain the 
wave package at the next time step $\psi^{n+1}=[{\bf T}^{n+1}]^{-1}\cdot \mathcal{V}^n$. 
The fractions of charmonium eigenstates 
are obtained by projecting the wave package to 
the wave function of the eigenstate.

\begin{figure}[!hbt]
\centering
\includegraphics[width=0.47\textwidth]{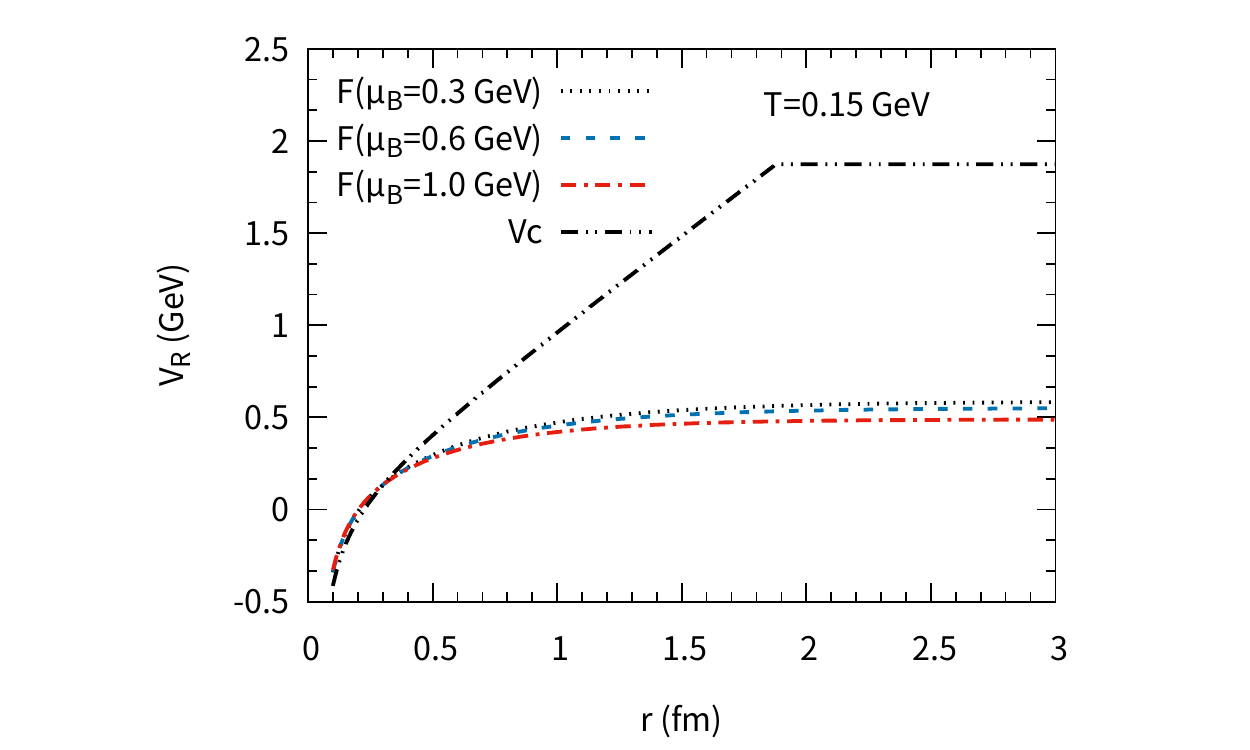}
\caption{ (Color online) 
The heavy quark potential as a function 
of radius at different temperatures. 
Dotted, dashed, dotted-dashed lines are the in-medium heavy quark potential (taken as the free energy F) with the baryon chemical potential $\mu_B=0.3, 0.6, 1.0$ GeV, respectively. The temperature is taken as $T=0.15$ GeV.  The Cornell potential is also plotted and labeled with $V_c$. 
}
\hspace{-0.1mm}
\label{lab-fig-vc-mu}
\end{figure}

The realistic in-medium heavy quark potential is between the limits of the free energy $F$ and the internal energy $U$. 
There are theoretical studies indicating that the in-medium potential is more close to the limit of $U$ in the temperatures available in AA collisions at RHIC and LHC~\cite{Liu:2010ej,Zhou:2014kka}. Consider that the internal 
energy $U=F+T(-\partial F/\partial T)$ can become a bit stronger than the vacuum Cornell potential at the temperatures around $T_c$~\cite{Islam:2020bnp,Wen:2022utn} which results in a oscillation behavior in 
the time evolution of charmonium fractions in the wave package, we take the free energy as the heavy quark potential to evolve the wave package.   
The real part of the potential is then parametrized with the form~\cite{Islam:2020bnp}, 
\begin{align}
\label{eq-realV}
    V_R(r,T,\mu_B)=
    -{\alpha\over r}e^{-m_d r} + {\sigma\over m_{d}}(1-e^{-m_d r}) 
\end{align}
where the Debye mass $m_d(T, \mu_B)$ depends on the temperature and 
the baryon chemical potential $\mu_B$~\cite{Doring:2005ih},
\begin{align}
    m_d(T,\mu_B)=  T
    \sqrt{{4\pi N_c\over 3}\alpha (1+{N_f\over 6})}  
    \sqrt{1+{3N_f\over (2N_c+N_f)\pi^2}
         ({\mu_B\over 3T})^2}
\end{align}
where the factors of color and flavor are taken as $N_c=N_f=3$. 
As we focus on the effect of baryon chemical potential $\mu_B$ at the collision energies of BES, the 
value of baryon chemical potential 
is estimated with the relation~\cite{Kadeer:2005aq,Li:2017ple},
\begin{align}
    \mu_B(\sqrt{s_{NN}})={1.3\over 1+0.28\sqrt{s_{NN}}}
\end{align}
In order to estimate the value of $\mu_B$ at the experiments of BES, we choose $\sqrt{s_{NN}}=10$ GeV to get a value of 
baryon chemical potential 
{$\mu_B\approx 0.3$ GeV. The value of $\mu_B$ can be 
larger than the medium temperature in the collisions of BES. The Debye mass is increased by the term with $\mu_B/(3 T)$. In the following calculations, we take different values of $\mu_B$ to check the $\mu_B$ effect. The color screened potential at finite $\mu_B$ is plotted in Fig.\ref{lab-fig-vc-mu}. }

\begin{figure}[!hbt]
\centering
\includegraphics[width=0.47\textwidth]{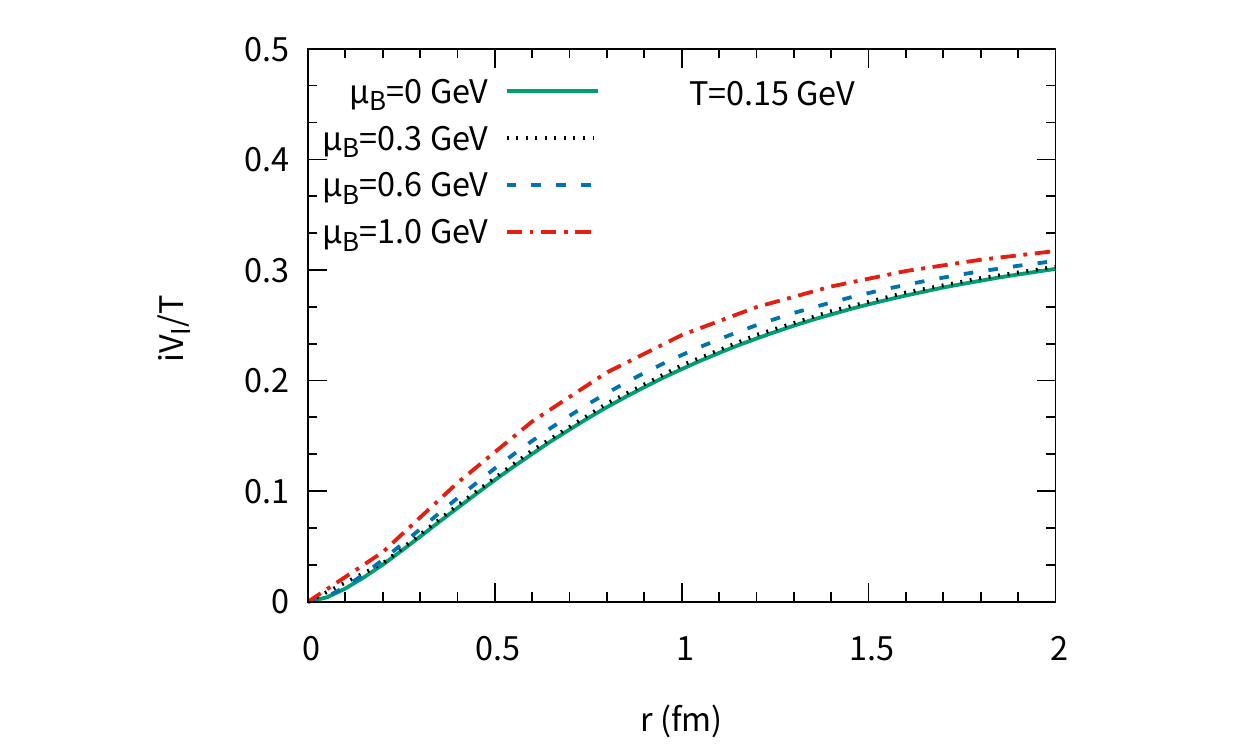}
\caption{ (Color online) 
The imaginary part of the potential scaled with the temperature $iV_I/T$ as a function of radius. Dotted, dashed, dotted-dashed lines with different values of $\mu_B$=(0,0.3,0.6,1.0) GeV are plotted respectively. The temperature is taken as $T=0.15$ GeV. 
}
\hspace{-0.1mm}
\label{lab-fig-pot-imag}
\end{figure}

Random inelastic scatterings with thermal partons can also dissociate quarkonium bound states in the medium which contributes an imaginary part in the potential of the singlet states. We take the parametrization based on the calculation from Hard Thermal Loop resummed perturbation theory~\cite{Laine:2006ns,Dumitru:2009fy},   
\begin{align}
\label{eq-imagV}
&V_I(r,T,\mu_B)=-i{g^2C_FT\over 4\pi} \tilde{f}(\hat{r})\\
&\tilde{f}(\hat{r})
=2\int_{0}^{\infty} dz {z\over (z^2+1)^2}
[1-{\sin(z\hat{r})\over z\hat{r}}]
\end{align}
{where $i$ is the imaginary unit. $C_F=(N_c^2-1)/(2N_c)$. 
$\hat r\equiv rm_d(T,\mu_B)$ is the dimensionless 
variable. The coupling constant is $g=\sqrt{4\pi \alpha N_c/3}$. The value of $\alpha$ is taken as the same with the Cornell potential. With this form, the
$\mu_B$ effect in $V_I$ is included via the Deybe mass. 
The magnitude of $V_I/T$ with 
different values of $\mu_B$ is plotted in Fig.\ref{lab-fig-pot-imag}.}

\section{Numerical results}
\begin{figure}[!hbt]
\centering
\includegraphics[width=0.47\textwidth]{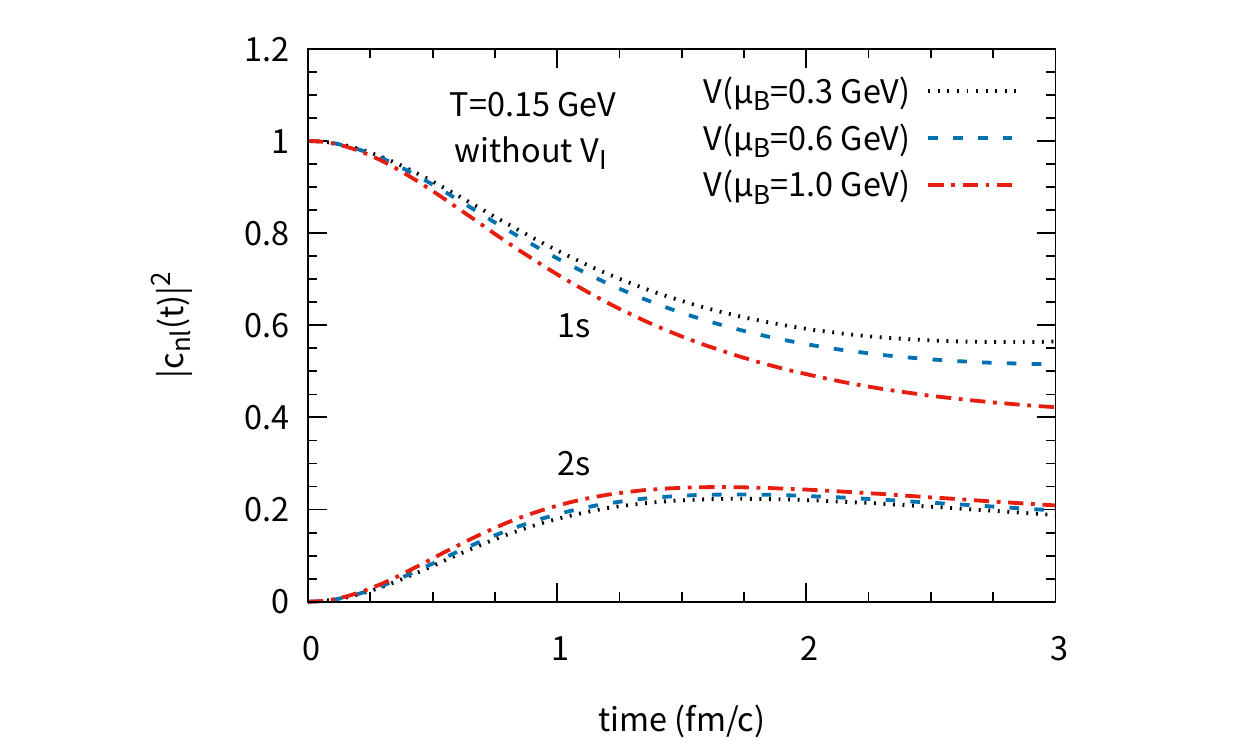}
\caption{ (Color online) 
The fraction of $J/\psi$ 
eigenstate in the wave package as  a function of 
time. The baryon chemical potential is taken 
as $\mu_B=0.3,0.6,1.0$ GeV respectively. Only real part of the potential is included in the calculations. 
The 
temperature of the static medium is $T=0.15$ GeV.}
\hspace{-0.1mm}
\label{lab-fig-VR}
\end{figure}

To study the effects of baryon chemical potential on the evolution of charmonium wave package, we take different values of $\mu_B$ in the calculations. 
The initial wave package is 
initialized with the wave function of $J/\psi$. 
In Fig.\ref{lab-fig-VR}, the temperature of the 
static uniformly-distributed medium is 
$T=0.15$ GeV. With 
only real part of the potential in Fig.\ref{lab-fig-VR}, the wave package 
expands outside, which reduces the quantum overlap 
between charmonium wave package and the wave 
function of $J/\psi$ state. As the geometry size 
of the excited state $\psi(2S)$ is larger than 
the size of $J/\psi$ wave function, the quantum overlap between the wave package and the $\psi(2S)$ 
wave function increases with time,  
shown as the lines in Fig.\ref{lab-fig-VR}. This behavior 
corresponds to the transitions of $J/\psi$ to 
$\psi(2S)$ components in the wave package. 
The Deybe mass with the baryon chemical potential $\mu_B=0.6$ GeV increases about $9\%$ compared 
with the case of $\mu_B=0$.
At high temperatures, 
the corrections 
of the $\mu_B$-term in the 
heavy quark potential 
become smaller. Time evolutions of $J/\psi$ fraction in the wave package
are close to each other when taking 
different values of $\mu_B$. 
In Fig.\ref{lab-fig-VR}, 
the sum of the fractions 
of $J/\psi$ and $\psi(2S)$ states 
become smaller than 
1, as some components 
of the wave package transform into higher 
eigenstates and scattering states due to 
the weak attraction in the wave package.

\begin{figure}[!hbt]
\centering
\includegraphics[width=0.47\textwidth]{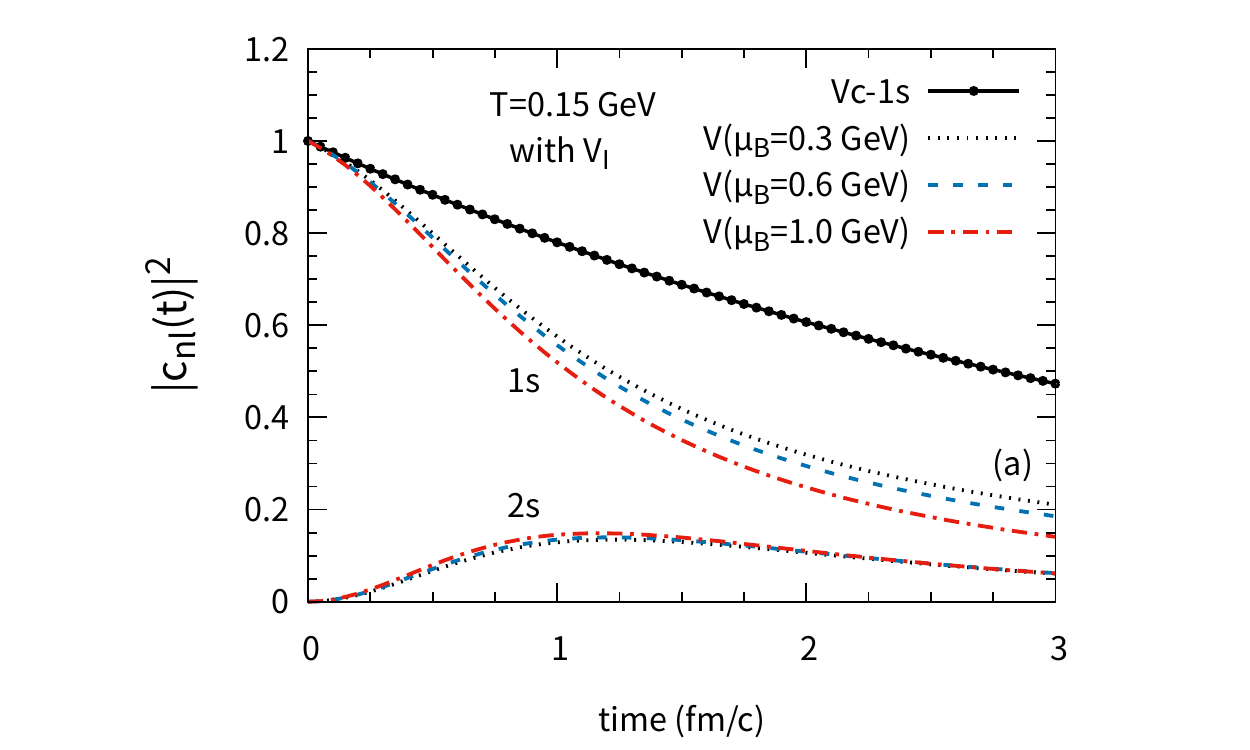}
\includegraphics[width=0.47\textwidth]{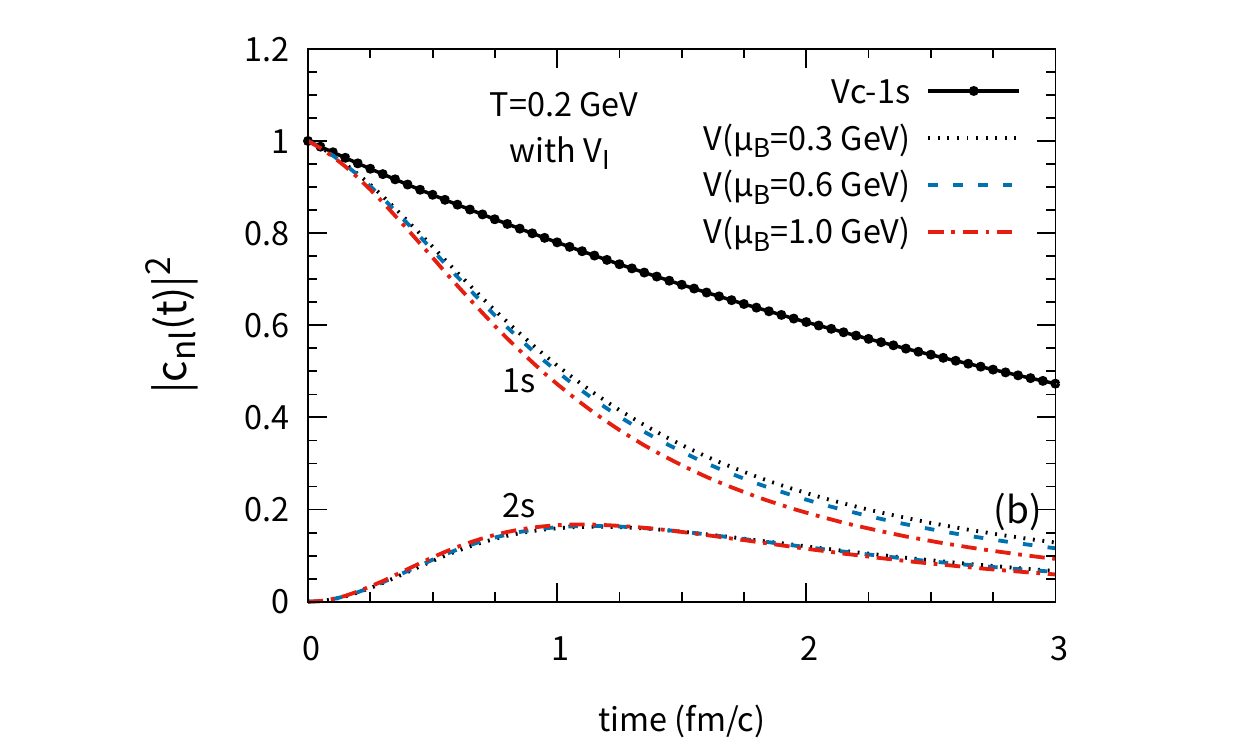}
\caption{ (Color online) The fraction of $J/\psi$ and $\psi(2S)$
eigenstate in the wave package as  a function of 
time. Both real and imaginary parts of the heavy 
quark potential are employed. {Medium temperature is taken as T=0.15 GeV and 0.2 GeV respectively, see the subfigure (a) and (b).} 
Black solid lines with markers employ the vacuum Cornell potential plus the 
imaginary potential. 
Other parameters are the same with Fig.\ref{lab-fig-VR}.
}
\hspace{-0.1mm}
\label{lab-fig-VR-VI}
\end{figure}

As introduced before, the transition from singlet to 
octet states induced by the parton random scatterings contributes an imaginary part 
in the potential of the singlet states. This 
reduces the normalization of the 
total wave package. 
After considering the imaginary potential 
given by Eq.(\ref{eq-imagV}), 
we study the 
$J/\psi$ survival probability in the static  
medium 
in Fig.\ref{lab-fig-VR-VI}. All the hot medium 
effects including color screening, 
$\mu_B$-correction, and inelastic scatterings are 
included. To check the contribution of the imaginary potential, we take the heavy quark potential to be 
$V_c+V_I(\mu_B=0)$, 
the reduction 
of $J/\psi$ fraction in the wave package is induced 
by the imaginary potential,  shown as the black 
solid lines with markers in Fig.\ref{lab-fig-VR-VI}. When the screened potential is also employed, $J/\psi$ 
fraction is more suppressed. At the time $t\sim 3$ fm/c, $J/\psi$ fraction with $\mu_B=1.0$ GeV 
is suppressed by around 25\% 
compared with the situation of $\mu_B=0$ at $T=0.15$ GeV. 
This effect becomes smaller at a higher temperature $T=0.2$ GeV, shown as the lower pannel of Fig.\ref{lab-fig-VR-VI}. 
In the long time limit, all the bound states will be dissociated by 
parton scatterings where the fractions of $J/\psi$ and $\psi(2S)$ go to zero.

In the relativistic heavy-ion collisions, 
hot medium is produced followed by a violent expansion. The medium temperature decreases with 
time. As a preliminary study, we neglect the transverse expansion of the medium and only consider the longitudinal expansion, where the temperature evolution can be characterized with the Bjorken model, 
\begin{align}
    {T(t)\over T(t_0)}=({t_0\over t})^{1/3}
\end{align}
where $t_0$ is the starting time of the bjorken 
expansion. From hydrodynamic models, it is 
estimated to be $t_0=0.6$ fm/c~\cite{Shen:2012vn,Hirano:2001eu}. The initial temperature is chosen as $T(t_0)=1.2\ T_c$, which is close to the initial temperature of the medium produced in BES collisions. The Schr\"odinger 
equation evolves until the temperature become 
lower than a cut $T_{f}=0.8\ T_c$ which is around the temperature of the medium kinetic freeze-out. Below this cut, heavy quark potential is taken as the vacuum Cornell 
potential.  

\begin{figure}[!hbt]
\centering
\includegraphics[width=0.47\textwidth]{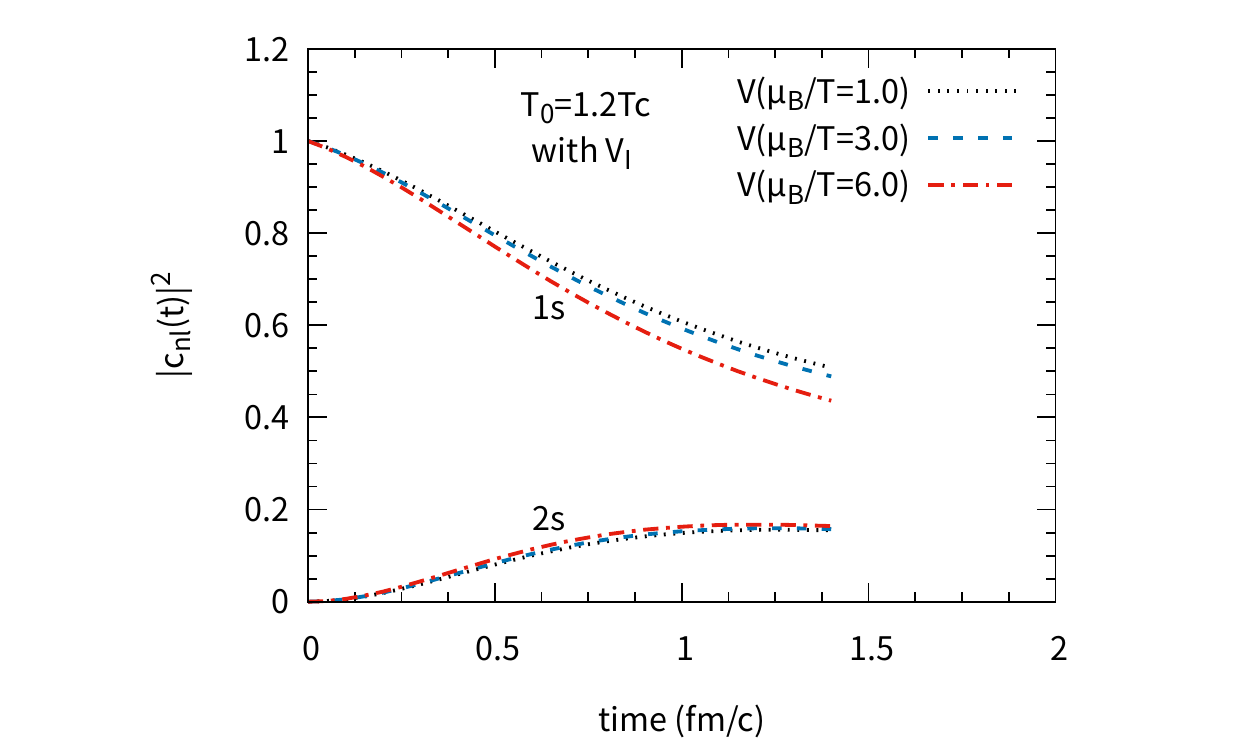}
\caption{ (Color online) The 
fraction of $J/\psi$ state in the 
wave package as a function of time in 
the Bjorken medium. 
Initial temperature of the 
medium is chosen as $T(t_0)=1.2\ T_c$. The 
starting time of the evolution is $t_0=0.6$ fm/c. 
Dotted, dashed, dotted-dashed lines correspond to the 
cases of  the complex potentials $V=F+V_I$ with 
$\mu_B/T=(1.0,3.0,6.0)$ respectively.}
\hspace{-0.1mm}
\label{lab-fig-bjork-VR-VI}
\end{figure}

In Fig.\ref{lab-fig-bjork-VR-VI}, the complex heavy quark potential is taken as the free energy plus the imaginary potential. 
{ In order to fix the value of entropy per baryon density, we take the value of $\mu_B/T=(1.0, 3.0, 6.0)$ respectively. The Deybe mass in both real and imaginary parts of the potential 
depends on $\mu_B/T$. In Fig.\ref{lab-fig-bjork-VR-VI}, one can see that the $J/\psi$ fraction is reduced by around $15\%$ in the line with $\mu_B/T=6.0$ compared with the situation of $\mu_B=0$ at the end of the bjorken medium evolution. $\mu_B$ effect can evidently reduce the charmonium survival probability in the baryon-rich medium. 
Note that in the dense medium, there is also friedel oscillation in the real part of the potential~\cite{Kapusta:1988fi}, which may also affect the evolution of quarkonium wave package. This effect is neglected in the work and deserves further studies in the future.
}

\section{Summary}
In this work, we employ the Schr\"odinger equation 
to study the evolutions of 
charmonium wave package at finite baryon chemical 
potential. $J/\psi$ fractions in the wave package 
is obtained by calculating the quantum overlap between the wave package and the wave function of $J/\psi$ eigenstate.  $\mu_B$ correction is included in the Deybe mass which is employed in both real and imaginary parts of the potential. With a large value of $\mu_B/T$, $J/\psi$ dissociation rate is enhanced in the baryon-rich medium. 
In the following work, we will also consider the $\mu_B$ dependence in the equation of state of the hot medium consistently.

\vspace{0.5cm}
{\bf Acknowledge:}
This work is supported by the
National Natural Science Foundation of China (NSFC)
under Grant Nos. 12175165, 11705125.

\end{document}